\documentclass[aps,prd,superscriptaddress,twocolumn,floatfix,letterpaper,nofootinbib]{revtex4-2}

\usepackage{xcolor}
\usepackage{graphicx}
\usepackage{natbib}
\usepackage{dcolumn}
\usepackage{bm}
\usepackage{hyperref}
\usepackage{amsmath}
\usepackage{bm}
\usepackage{booktabs}

\begin{document}

\title{Constraining Neutrino Mass with the Void Weak Lensing Effect}

\author{Wenshuo Xu}
\email{xws21@mails.tsinghua.edu.cn}
\affiliation{Department of Astronomy, Tsinghua University, Beijing 100084, China}
\author{Cheng Zhao}
\email{czhao@tsinghua.edu.cn}
\affiliation{Department of Astronomy, Tsinghua University, Beijing 100084, China}
\author{Chen Su}
\affiliation{Shanghai Astronomical Observatory (SHAO), Nandan Road 80, Shanghai 200030, China}
\affiliation{University of Chinese Academy of Sciences, Beijing 100049, China}
\author{Huanyuan Shan}
\affiliation{Shanghai Astronomical Observatory (SHAO), Nandan Road 80, Shanghai 200030, China}
\affiliation{University of Chinese Academy of Sciences, Beijing 100049, China}
\author{Yu Liu}
\affiliation{Department of Astronomy, Tsinghua University, Beijing 100084, China}
\affiliation{Institute of Physics, Laboratory of Astrophysics, École Polytechnique Fédérale de Lausanne (EPFL), Observatoire de Sauverny, CH-1290 Versoix, Switzerland}

\date{\today}

\begin{abstract}
Cosmic voids, the underdense regions of the Large Scale Structure (LSS), provide cosmological information highly complementary to that obtained from overdense regions. In this work, we investigate the constraining power of the void-shear cross-correlation (void lensing effect) on the total neutrino mass. Based on cosmological $N$-body simulations with varying neutrino masses, we populate BOSS LOW-Z-like galaxies at $0.2<z<0.4$ using HOD fitting, identify voids with the {\sc dive} void finder and obtain their density profiles from the underlying dark matter and neutrino distributions. We then generate mock shear catalogues through ray-tracing and measure the corresponding void lensing signals, assuming a source number density of $10/{\rm arcmin}^{2}$ and sky area of around $8400\,{\rm deg}^2$. Under this setup, void lensing independently yields a constraint on total neutrino mass as $\sigma(M_{\nu})=0.096\,{\rm eV}$ ($M_{\nu}<0.232\,{\rm eV}$, 95\% C.L.) in the absence of shape noise, and $\sigma(M_{\nu})=0.340\,{\rm eV}$ ($M_{\nu}<0.707\,{\rm eV}$, 95\% C.L.) when adopting a Stage-III-like shape noise. Moreover, we find a clear linear relationship between the void lensing signal and neutrino mass. We further validate the forward modelling of the void lensing signal from the void density profiles across different cosmologies, demonstrating its accuracy and potential for future applications. These findings highlight void lensing as a promising probe of massive neutrinos and motivate its applications to galaxy survey data as well as the combination with other cosmological observables.

\end{abstract}

\maketitle

\section{Introduction}
\label{sec:introduction}

As a standard cosmological paradigm, the $\Lambda$CDM model succeeds in fitting a number of independent observations. Experiments on the Cosmic Microwave Background (CMB) anisotropies provide tight constraints on the six $\Lambda$CDM parameters \cite{Planck2018, ACT_DR6, SPT_result}, while other low-redshift probes make further improvements by breaking parameter degeneracies.
However, there are still remaining open questions for the $\Lambda$CDM model (e.g. Ref.~\cite{cosmological_tension}). Meanwhile, with the increasing volume and tracer density of cosmological surveys, more attention has been devoted to extensions of the vanilla $\Lambda$CDM model, including dynamical dark energy (e.g. Ref.~\cite{Lodha2025}), modified gravity (e.g. Ref.~\cite{Ishak2025}), primordial non-Gaussianity (e.g. Ref.~\cite{Chaussidon2025}) and massive neutrinos (e.g. Ref.~\cite{Elbers2025}).

The state-of-the-art constraints on total neutrino mass ($M_{\nu}=\Sigma\,m_{\nu}$) arise from the combination of particle physics experiments and cosmological observations. The recent KATRIN experiment places an upper limit of $M_{\nu}<1.35\, {\rm eV}$ (90\% C.L.) when combined with neutrino oscillations \cite{KATRIN_Collaboration}, while oscillation experiments further set lower limits for the two possible mass orderings \cite{particle_physics_review}. In parallel, cosmological constraints on neutrino mass have also steadily tightened (e.g. Refs.~\cite{Vagnozzi2017, Choudhury2020, Brieden2022, Elbers2025}). As massive neutrinos remain relativistic until relatively late time and further suppress the growth of small-scale density perturbations through free streaming, they leave imprints on both the cosmic expansion history and large-scale matter distribution. Using Baryon Acoustic Oscillations (BAO) measurements to constrain the expansion history, the Dark Energy Spectroscopic Instrument (DESI) Collaboration has reported the tightest constraint to date as $M_{\nu}<0.0642\,{\rm eV}$ (95\% C.L.) when combined with CMB data \cite{DESI_DR2_BAO, Elbers2025}. However, this result is known to be affected by prior weight effects, with the data favoring negative values of the effective neutrino masses \cite{effective_Mnu}. Such tension implies potential unknown systematics or new physics, such as dynamical dark energy \cite{Elbers2025}. While these constraints all rely on the cosmic expansion history, neutrino free-streaming effect can provide complementary information, which is manifested as a scale-dependent suppression of the power spectrum \cite{Hu1998, Abazajian2016, DESI_DR1_FS}. Beyond the 2-point clustering of galaxies, several higher-order probes sensitive to neutrino free-streaming have also been proposed to constrain massive neutrinos, such as the bispectrum \cite{Belsunce2019, Kamalinejad2020}, Minkowski functional \cite{min_functional, Liu2023}, persistent homology \cite{Yip2024, Kanafi2024, betti_curve_2025} and cosmic voids \cite{Massara2015, void_cmblensing, Thiele2024}.

Cosmic voids, which trace underdense regions of the LSS and constitute most of the volume in the Universe, have been shown to provide complementary cosmological information to that from overdense regions \cite{Joeveer1978, Gregory1978, Sutter2012, Pan2012, Libeskind2018, Douglass2023}. Unlike gravitationally collapsed structures, voids still mostly maintain linear evolution features and are marginally affected by baryonic effects \cite{Hamaus2014, Paillas2017, Schuster2024}. This suggests voids to be powerful probes of several cosmological effects, such as dark energy \cite{Pisani2015, eBOSS_void_BAO}, modified gravity \cite{Perico2019, Su2023}, primordial non-Gaussianity \cite{Kamionkowski2009} and massive neutrinos \cite{Massara2015}.

Owing to their large thermal velocities, massive neutrinos are less clustered than cold dark matter, leading to a relatively higher neutrino fraction in underdense regions.
This makes voids a promising environment to constrain neutrino properties (e.g. Refs.~\cite{Massara2015, Kreisch2019, Schuster2019, Thiele2024, void_cmblensing}). Ref.~\cite{Massara2015} performed the first systematic study of cosmic voids in massive neutrino cosmologies and found that, with the presence of massive neutrinos, voids tend to present larger sizes, higher ellipticities, less empty inner densities and smaller radial velocity magnitudes. They also showed that these differences become less pronounced at higher redshifts, as neutrinos are more relativistic and less affect the structure formation. In particular, based on these results, the authors proposed the void density profile as a sensitive probe of neutrino mass and further discussed the challenges of theoretical modelling in Ref.~\cite{Massara2018}.

From the observational perspective, an additional challenge in using void density profiles for cosmological constraints arises from that the distributions of dark matter and neutrinos inside voids cannot be directly observed.
Several works (e.g. Refs.~\cite{Nadathur2016, Montero-Dorta2025}) have tried to use galaxies as biased tracers of underlying matter distribution but require careful modelling of galaxy bias, which is complicated in void environment \cite{Montero-Dorta2025, Alfaro2026}. In contrast, gravitational lensing, as a purely gravitational effect, provides a direct probe of the total matter distribution \cite{grav_lens_book}.
In the weak lensing regime, lensing shear measurements enable to both constrain the matter distribution around specific foreground lens objects and probe the large-scale matter distribution in the Universe \cite{kids_3x2pt, DES_3x2pt, desi_3x2pt}. Following this idea, the weak lensing effect of cosmic voids facilitates the measurement of void density profile. 

The weak lensing effect of cosmic voids has been extensively studied. 
Observationally, measurements have been made in both photometric and spectroscopic surveys, and recently Ref.~\cite{Martin2025} reported the highest-significance detection to date of a $6.2\sigma$-measurement using spectroscopically identified voids in SDSS-III BOSS and UNIONS lensing shear catalogues. On the theoretical side, void lensing has been shown to be sensitive to several cosmological effects, such as modifications of gravity \cite{Barreira2015, Baker2018, Su2023} and massive neutrinos \cite{void_cmblensing, Maggiore2025}, owing to the reduced screening effect and high neutrino mass fraction in underdense environments. However, the constraining power of void lensing on neutrino mass is still largely unexplored.

In practice, void studies are further complicated by the emergence of multiple void-finding algorithms, owing to the absence of an explicit definition for void spatial boundaries. Most algorithms rely on triangulation or tessellation (e.g. Refs.~\cite{wvf_finder, zobov, ocvf_finder}), while some others adopt numerical estimation of the local density contrast \cite{svf_finder}. Ref.~\cite{Davies2018} also proposed the direct construction of voids in lensing convergence maps. As voids defined in different ways can exhibit different properties \cite{void_comparion_1, void_comparison_2}, it is important to choose the definition providing adequate information for specific studies.

In this work, we adopt the {\sc dive} void finder developed in Ref.~\cite{Zhao2016}. The resulting Delaunay Triangulation (DT) voids have been shown to achieve a high statistical S/N ratio \cite{Kitaura_void_BAO, eBOSS_void_BAO}, which benefits our void lensing analysis. 
Using $N$-body simulations with varying total neutrino masses, we generate realistic mock galaxy and shear catalogues, and investigate the constraining power of DT-void lensing effect on the neutrino mass. We also validate the modelling of void lensing signal based on the void density profile. With this model and the covariance estimated from multiple realizations, we provide a quantitative forecast through posterior sampling. For simplicity, we fix all cosmological parameters other than $M_{\nu}$, neglect redshift-space distortions (RSD) and identify voids in real space.

This paper is structured as follows: We present the basic formalism of weak lensing in Section~\ref{sec:lensing_formalism}, and introduce the simulation data and methods used to generate mocks in Section~\ref{sec:data}. Then Section~\ref{sec:void_measurement} shows the measurements of void density profiles and weak lensing signals from our simulation and mock data. Finally, we provide the inference on neutrino mass in Section~\ref{sec:constraint}, and discuss our results and conclude in Section~\ref{sec:summary}.

\section{Weak Lensing Formalism}
\label{sec:lensing_formalism}

In this section, we briefly review the formalism of weak gravitational lensing and the derived equations related to this work. More comprehensive reviews can be found in Refs.\cite{grav_lens_book} and \cite{bartelmann_2017}.

When light from distant galaxies propagates through the foreground large-scale structures, its path is perturbed by the gravitational field.
In the weak lensing regime, this deflection effect induces distortions in the observed shape of distant galaxies, which can be described by the Jacobian matrix:
\begin{equation}
    \label{eq:jacobian}
    A=\begin{pmatrix} 1-\kappa-\gamma_1 & -\gamma_2 \\ -\gamma_2 & 1-\kappa+\gamma_1 \end{pmatrix}    
\end{equation}
, where the \textit{convergence} $\kappa$ indicates the overall size magnification, while the \textit{shear} $\gamma_1$ and $\gamma_2$ describe the change in ellipticity.

When isolating the weak lensing effect of an axially symmetric foreground lens (such as DT voids in this work; see Section~\ref{subsubsec:dive}), its convergence profile can be expressed in thin-lens approximation as
\begin{equation}
\label{eq:kappa_int}
    \kappa(R)=\frac{\bar{\rho}_{m,0}}{\Sigma_{c, {\rm comov}}}\,\int\delta(r=\sqrt{R^2+z^2})\,dz,
\end{equation}
where $R$ is the 2D projected radius, $\delta$ is the overdensity and $\Sigma_{c, {\rm comov}}=\frac{c^2}{4\pi G}\left[ \frac{\chi_s\,a_l}{(\chi_s-\chi_l)\chi_l}\right]$ is the \textit{comoving} critical surface mass density. Here the subscripts $s$ and $l$ stand for the source and lens, respectively, and $\chi$ is the comoving distance.
The shear effect around the foreground lens can be further transferred to the tangential ($\gamma_t$, E-mode) and cross ($\gamma_{\times}$, B-mode) terms, and the tangential component is approximated by the convergence as

\begin{equation}
    \gamma_t(R)=\bar{\kappa}(<R)-\kappa(R),
\end{equation}

where $\bar{\kappa}$ is the convergence averaged within a given radius, $\bar{\kappa}(<R)=2/R^2\int_0^{R}r\kappa(r)\,dr$.
A generally chosen measurement in observations is the excess surface mass density (ESD) $\Delta\Sigma$, which is defined as
\begin{equation}
\label{eq:esd}
    \Delta\Sigma(R)=\Sigma_{c, {\rm phys}}\gamma_t=\Sigma_{c, {\rm phys}}(\bar{\kappa}-\kappa),
\end{equation}

where the \textit{physical} critical surface mass density $\Sigma_{c, {\rm phys}}=\frac{c^2}{4\pi G}\,\frac{D_s}{D_{ls}D_l}=\Sigma_{c, {\rm comov}}/{a_l^2}$, $D$ represents angular diameter distance and $a_l$ is the scale factor at the redshift of foreground lens. 
$\Delta\Sigma$ reflects the projected mass distribution around the lens objects and maximizes the S/N ratio of measurement by weighting sources at different redshifts with critical surface denisty \cite{Shirasaki2018}.

Eqs.~(\ref{eq:kappa_int})-(\ref{eq:esd}) clarify the relations between lens density profile and weak lensing signal, which indicates we can constrain the density profile through weak lensing measurements, as well as theoretically forward model the lensing signal from density profile. When processing simulation and mock data, density profile can be measured through the cross-correlation between voids and dark matter and we adopt the estimator of $\Delta\Sigma$
\begin{equation}
\label{eq:esd_estimator}
    \widehat{\Delta\Sigma}=\frac{\Sigma_{i,j}w_{ij}e_{t, ij}\Sigma_{c,ij}}{\Sigma_{ij}w_{ij}},
\end{equation} 
where $i$ and $j$ go over all of the foreground lens objects and background sources, respectively, and the tangential ellipticity of source galaxies $e_t$ works as an unbiased estimator of the lensing tangential shear $\gamma_t$ in Eq.~(\ref{eq:esd}). The weight function $w_{ij}$ is assigned to each lens-source pair to improve the statistical S/N in observational studies \cite{Sheldon2004, Mandelbaum2005}. In this simulation-based work, we set $w_{i,j}=1.0$ for simplicity.

\begin{figure}
\centering
\includegraphics[width=0.45\textwidth]{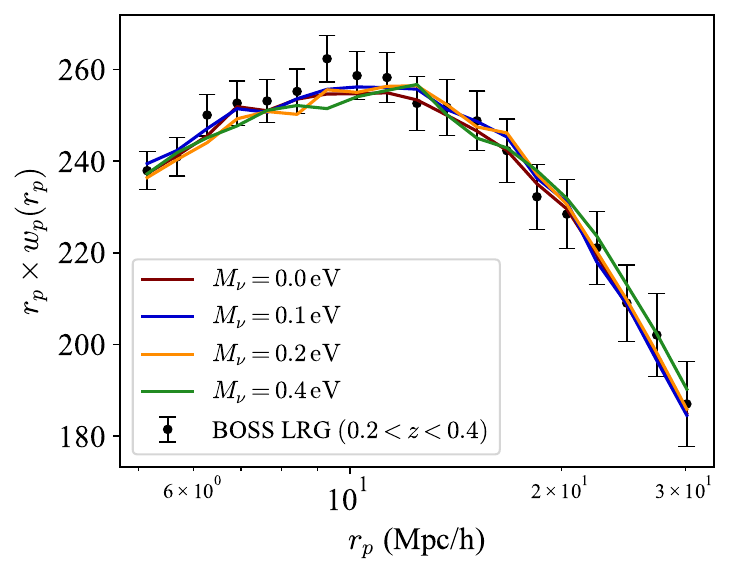}
\caption{The results of HOD fitting. Lines in different colors indicate the best-fit projected 2-point correlation function in simulations with different neutrino masses. The black data points show the projected clustering of BOSS LRGs in $0.2<z<0.4$ and the errorbars are estimated with 1000 PATCHY mocks.}
\label{Fig_HOD}
\end{figure}

\section{Simulations and Mocks}
\label{sec:data}
In this section, we present the data and methods used to generate mocks. We first introduce the setup of our $N$-body simulations with different neutrino masses in Section~\ref{subsec:nbody}, then discuss the mock DT void catalogues in Section~\ref{subsec:void}. Finally, Section~\ref{subsec:raytracing} presents our raytracing pipeline and the incorporation of observational effects.

{\subsection{$N$-body Simulations with Massive Neutrinos}
\label{subsec:nbody}

To investigate the impact of massive neutrinos on the void lensing signal, we employ four $N$-body simulations with different total neutrino masses. We use the $M_{\nu}=0.0$ and $0.1\,{\rm eV}$ realizations from \textit{Jiutian} extension runs \cite{Jiutian}, and generate two additional realizations with $M_{\nu}=0.2$ and $0.4\,{\rm eV}$ in the same manner using the publicly available information-optimized code {\sc cube} \cite{CUBE_code1, CUBE_code2}.}

These simulations co-evolve cold dark matter and neutrinos in a cubic box of side length $1500\, {\rm Mpc/h}$ with periodic boundary conditions, using $N_{\rm part}=2048^3$ dark matter particles. To avoid the shot noise problem of neutrinos, we adopt grid-based method developed in Ref.~\cite{CUBE_code2} and evolve massive neutrinos on a mesh with $N_{\rm pix}=2048^3$.

For the two additional runs, we use the same initial Gaussian random field as the two from Jiutian to cancel the impact of cosmic variance on our analyses, and adopt \textit{Planck2018} cosmology \cite{Planck2018}: $\Omega_m=0.3111$, $h=0.6766$, $n_s=0.9665$ and $A_s=2.100549\times10^{-9}$. For simulations with massive neutrinos, we include a non-zero $\Omega_{\nu}$ and reduce $\Omega_{\rm cdm}$ to make the total matter density invariant. Then the initial power spectrum and transfer functions are recalculated with {\sc camb} \cite{camb_paper}.

\subsection{Void Catalogue}
\label{subsec:void}

In this work, we simulate voids identified from galaxy 3D positions, which can be measured in spectroscopic surveys. For simplicity, we only study voids around $z_l=0.3$.

From $N$-body simulations, we adopt the {\sc rockstar} halo finder \cite{rockstar_paper} to generate catalogues of halos with more than 20 dark matter particles, corresponding to a lower limit of halo mass around $6.8\times10^{11}\,{\rm M_{\odot}}/h$. Then we populate galaxies according to the clustering of BOSS LRG galaxies in $0.2<z<0.4$ through the vanilla HOD model (Section~\ref{subsubsec:hod}). Finally, we identify void catalogues from galaxies with the {\sc dive} void finder (Ref.~\cite{Zhao2016}, and briefly reviewed in Section~\ref{subsubsec:dive}).

\subsubsection{Halo Occupation Distribution (HOD)}
\label{subsubsec:hod}
In this work, we generate galaxy catalogues from dark matter halos 
with the vanilla five-parameter HOD model introduced in Ref.~\cite{zheng2007_HOD}. In HOD models, galaxies are separated into central and satellite galaxies. The occupation of central galaxies is assumed to follow a Bernoulli distribution described by $M_{\rm min}$ and $\sigma_{\log{M}}$, while satellites follow a Poisson distribution with $M_0$, $M_1$ and $\alpha$. We refer interested readers to the original paper for the specific formulae.

As we only focus on voids around $z=0.3$ in this work, we measure the projected 2-point correlation function $w_p(r_p)$ of SDSS-III BOSS LRG galaxies at $0.2<z<0.4$ (LOWZ samples in Ref.~\cite{BOSS_DR12}) and use it as the data vector in HOD fitting. The projected correlation function, which is the compressed redshift-space 2-point correlation function $\xi(r_p, r_{\pi})$, can be expressed as
\begin{equation}
\label{eq:w_p}
    w_p(r_p)=2\int_0^{r_{\pi, \rm max}}\xi(r_p,\,r_{\pi})dr_{\pi},
\end{equation}
where $r_p$ and $r_{\pi}$ are the transverse and light-of-sight separations, respectively. In this work, we use the Landy \& Szalay estimator \cite{Landy_Szalay}:
\begin{equation}
    \xi(r_p,\,r_{\pi})=\frac{DD-2DR+RR}{2},
\end{equation}
where $R$ is the random catalogue, and we calculate the integration in Eq.~\ref{eq:w_p} with $r_{\pi,\,\rm max}=35\,{\rm Mpc/h}$. The covariances used for HOD fitting are estimated with 1000 PATCHY mocks \cite{patchy_mock}. We present the HOD fitting results in Fig.~\ref{Fig_HOD}, which indicates a great fit in all four simulations. Then with the best-fit HOD parameters, we populate mock galaxies with {\sc halotools} \cite{halotools_paper}, and randomly downsample the four galaxy catalogues to the same number density $3\times10^{-4}\,{\rm Mpc}^{-3}h^3$ to keep aligned with BOSS galaxies. In addition, as the DT voids are defined in a geometric way, a higher tracer number density will systematically suppress the abundance of large voids, which can be avoided with the downsampling. We also further discuss the propagation of HOD uncertainties into void identification in Appendix \ref{appendix_d}.

\begin{figure}
\centering
\includegraphics[width=0.45\textwidth]{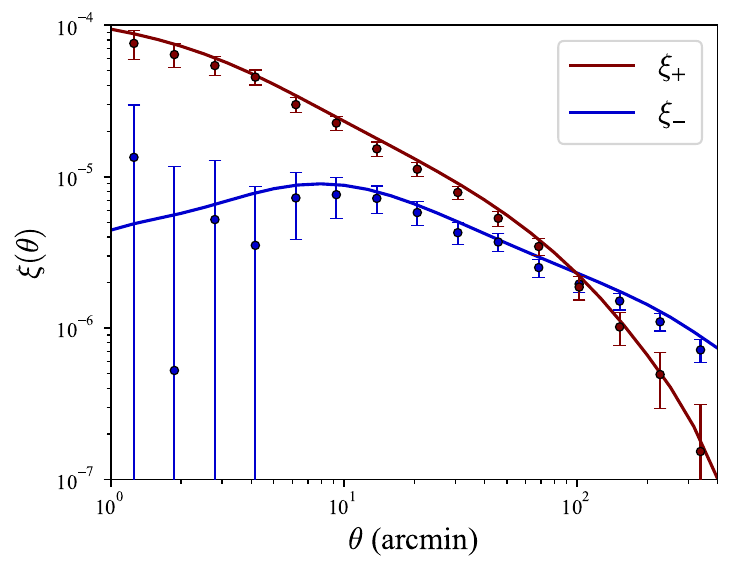}
\caption{Shear auto-correlation functions $\xi_{+}$ (red) and $\xi_{-}$ (blue). The data points are the measurements in our mock source catalogues using {\sc treecorr} \cite{treecorr} and the errorbars are estimated in 15 realizations (see Section~\ref{subsec:covariance}). Here we incorporate a shape noise $\sigma_e=0.3$ in the catalogues. The solid lines are the prediction of 3D matter power spectrum calculated with {\sc pyccl} \cite{pyccl}.}
\label{Fig_shear_corr}
\end{figure}

\subsubsection{Void Finder}
\label{subsubsec:dive}

In this work, we employ the Delaunay trIangulation Void findEr ({\sc dive}) developed in Ref.~\cite{Zhao2016}, which samples underdense regions using the circumspheres of the Delaunay Triangulation of tracers. The \textit{DT voids} identified with {\sc dive} show a large overlapping fraction, leading to a high number density. The clustering measurement of DT voids shows this high density results in an improved S/N ratio for statistical studies \cite{Kitaura_void_BAO, eBOSS_void_BAO}. We identify voids from the galaxy catalogues described in Section~\ref{subsubsec:hod} with {\sc dive} and use the original overlapping void catalogues in the analysis. Moreover, the spherical shape and naturally defined void radius of DT voids further facilitate the theoretical modelling of lensing signal through Eqs.~(\ref{eq:kappa_int})-(\ref{eq:esd}). We also apply a radius selection and only use voids within range $17<R_V<25\,{\rm Mpc/h}$, as small voids have been shown to trace overdense regions \cite{Sheth2004} and larger voids suffer from much lower number densities (Appendix~\ref{appendix_b}). The total number density of the selected voids is approximately $4.7\times10^{-4}\,{\rm Mpc^{-3}h^3}$, with slight variations among different cosmologies (Fig.~\ref{Fig_VSF}).

\begin{figure*}\centering
	\includegraphics[width=\textwidth]{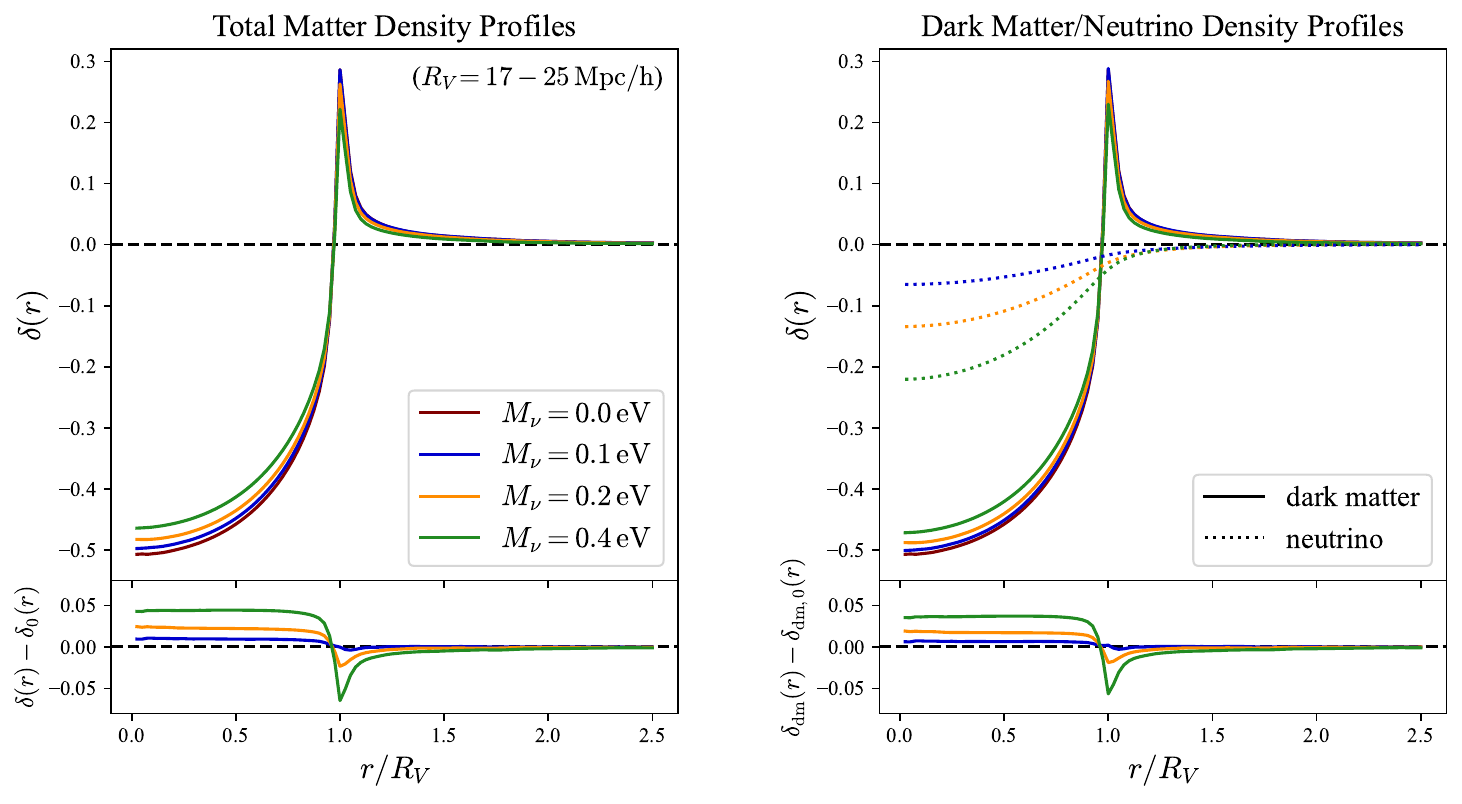}
    \caption{Combined density profiles of voids with $17<R_V<25$\,Mpc/h measured in simulations. Different colors represent simulations with different neutrino masses. The left panel shows the total density profile, while the right panel presents dark matter and neutrino components separately with solid and dotted lines. The bottom panels display the differences between simulations with massive neutrinos and massless neutrinos ($\delta_0(r)$.}
    \label{Fig_vprof}
\end{figure*}

\subsection{Ray-tracing Mocks}
\label{subsec:raytracing}

To further investigate the observational effects, as well as validate the theoretical modelling of void lensing signal described in Section~\ref{sec:lensing_formalism}, we generate mock shear catalogues following the multiplane ray-tracing method introduced in Ref.~\cite{Hilbert2009}.  To simplify the analysis, we place all source galaxies at $z_s=0.75$ and run ray-tracing up to this redshift with an angular resolution $\sim0.8\,{\rm arcmin}$. The lensing efficiency for $z_s=0.75$ peaks around $z_l=0.3$, which is consistent with our foreground galaxy and void catalogues (Section~\ref{subsubsec:hod}).
Following Ref.~\cite{Hilbert2009}, we adopt the flat-sky approximation, which should not bias our results given that we only focus on scales smaller than $50\,{\rm Mpc/h}$ (twice the maximum radius of our selected void samples).

Limited by the boxsize, it is unable to ray-trace to $z=0.75$ in a single simulation box. One possible solution is to replicate boxes to increase the volume (e.g. Ref~\cite{Su2023}), but may suffer from artificial repetition of structures (see Ref.~\cite{Chen2024}, who also explore the strategies to minimize this impact by optimizing the line-of-sight selection). Ref.~\cite{Carlson2010} proposed an alternative remapping method to transform a cubic box into a rectangular cuboid, with preserving the continuity and avoiding the repetition of structures. We refer to this method to generate a suite of density slices with a thickness of $L \approx125\,{\rm Mpc/h}$ for ray-tracing (except for the first and last slices, see Appendix~\ref{appendix_c} for more details).
The redshift evolution of structures is incorporated by using snapshots saved with $\Delta z=0.05$. As the dark matter and neutrino components are realized in different ways in the simulations, we smooth the dark matter particles with a CIC kernel and also resample the original neutrino overdensity mesh to get the total matter distribution in each slice.

With the ray-tracing results at $z=0.75$, we randomly sample source galaxy catalogues with a number density $10/{\rm arcmin}^{2}$, which lies between the typical values of stage-III surveys \cite{kids_catalogue, des_catalogue} and stage-IV predictions \cite{Euclid_design, csst_euclid_synergy}.
The shear values of each source are linearly interpolated from ray-tracing results. We further incorporate a Gaussian random shape noise with $\sigma_e=0.3$ to mock real observations.

To obtain the consistent foreground void catalogues, we first remap galaxies in simulation box to a slice using the same method, and subsequently identify voids with the remapped galaxy positions. Since the periodic boundary conditions are partially broken by the remapping, we drop voids with centers outside the slice. The slice thickness $125\,{\rm Mpc/h}$ corresponds to a redshift width $\delta z\approx0.05$ at $z_l=0.3$, which is actually narrower than the typical redshift bins adopted in galaxy-galaxy lensing studies.

In addition, the transverse extent of the remapped slices allows for a split into 15 subregions with an angular area of $8400\,{\rm deg}^2$ at $z=0.3$ (see Appendix~\ref{appendix_c}), which is comparable to the overlap between DESI and \textit{Euclid} \cite{Naidoo2023}.
These subregions are used as multiple independent realizations for the covariance estimation in Section~\ref{subsec:covariance}. Fig.~\ref{Fig_shear_corr} shows the shear auto-correlation functions, $\xi_{+}$ and $\xi_{-}$, of our mock source catalogues averaged over all 15 sub-regions, demonstrating a good agreement between ray-tracing results and the prediction of 3D matter power spectrum. Finally, we note that, as a compromise for the limited simulation data, the flat-sky approximation and remapping method adopted in generating mocks may lead to slight inaccuracies in the covariance estimation. Given that we only focus on the cross-correlation in this work, and the covariance is dominated by the shape noise (Fig.~\ref{Fig_corr_matrix}), we do not expect these effects would bias our conclusions and dedicate more accurate shear mocks to future work.

\section{Void Statistics}
\label{sec:void_measurement}
In this section, we present the measurement of void density profile (Section~\ref{subsec:void_profile}) and void lensing signals (Section~\ref{subsec:void_esd}) from the simulation and mock data. We also compare the theoretical model of $\Delta \Sigma$ and measurements from mocks in Section~\ref{subsec:void_esd}. It is noted that the results shown in this section are the average of all voids with radius $17<R<25\,{\rm Mpc/h}$ (see Section~\ref{subsubsec:dive}), which is also used for the cosmological inference in next section. We present the results in separate radius bins in Appendix~\ref{appendix_a}.

\begin{figure}
\centering
\includegraphics[width=0.45\textwidth]{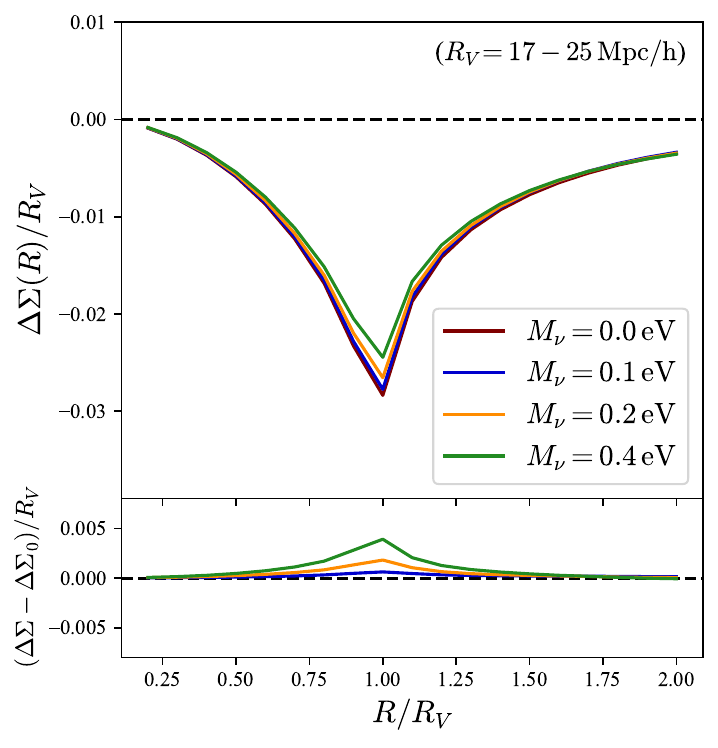}
\caption{Theoretical model of the combined void lensing signal $\Delta \Sigma$ for voids with $17<R<25$\,Mpc/h. Different colors indicate different neutrino masses. The bottom panel shows the differences between results of massive neutrinos and the massless case.}
\label{Fig_theo_ESD}
\end{figure}

\subsection{Void Density Profile $\delta(r)$}
\label{subsec:void_profile}

Fig.~\ref{Fig_vprof} shows the void density profiles measured in the $z=0.3$ snapshot of simulations with different neutrino masses. The results are estimated from the complete void catalogue within the simulation box, and exhibit negligible jackknife errors, with an average S/N ratio of order $10^3$, owing to the high number density of DT voids.

The results demonstrate that the selected DT voids trace underdense regions with the negative central density contrast. The left panel of Fig.~\ref{Fig_vprof} shows the total matter distribution. For DT voids, they always present a sharp density peak at boundary ($r=R_V$), which is distinguishable from other voids, such as the universal profile \cite{Hamaus2014} for voids identified with modified {\sc zobov}, which have much smoother density walls at boundary. The underlying reason for this feature is that, by definition, there are always four galaxies located at void boundary. As galaxies typically reside in high-density regions, this results in a significant density increase. At larger radial distance, the density returns to homogeneity ($\delta=0$) when averaging over a large sample of voids. 

Similar to the results presented in Ref.~\cite{Massara2015}, we find that for DT voids, massive neutrinos also lead to a higher density at the central underdense regions and a lower density peak at boundaries, with these changes increasing with neutrino mass.

To further demonstrate this behavior, we show the separate distributions of neutrinos and dark matter in the right panel of Fig.~\ref{Fig_vprof}, which indicates that neutrinos broadly trace the large-scale dark matter distribution, but present significantly weaker clustering. This behaviour arises from their large thermal velocities, which suppress gravitational collapse below the free-streaming scale. At fixed $\Omega_m$, the presence of massive neutrinos correspondingly reduces dark matter density. Because neutrino distribution remains much smoother than that of dark matter, their contribution to the gravitational potential cannot compensate for the reduced clustering of cold dark matter. Consequently, the evolution of large-scale structure slows down, leading to shallower void density profiles. Although more massive neutrinos cluster more efficiently due to their lower thermal velocities, their clustering is still subdominant on the scales of voids. Therefore, the combined effect, which is dominated by the reduction in $\Omega_{\rm cdm}$, results in progressively shallower void density profiles with increasing neutrino mass.

We clarify that, although the observed effect is driven by the reduction in $\Omega_{\rm cdm}$, its physical origin lies in the large thermal velocities of massive neutrinos, which suppress their clustering and prevent efficient contributions to the gravitational potential. 
However, we note that our result is subject to the degeneracy between $\Omega_m$ and neutrino mass, as demonstrated in many previous studies \cite{Elbers2025}. This degeneracy cannot be fully explored with our current simulation set and may result in an overestimation of the constraining power. Nevertheless, we expect that this degeneracy can be effectively broken by combining void lensing with other cosmological probes and leave a more detailed investigation for future work.

\begin{figure*}
\centering
\includegraphics[width=\textwidth]{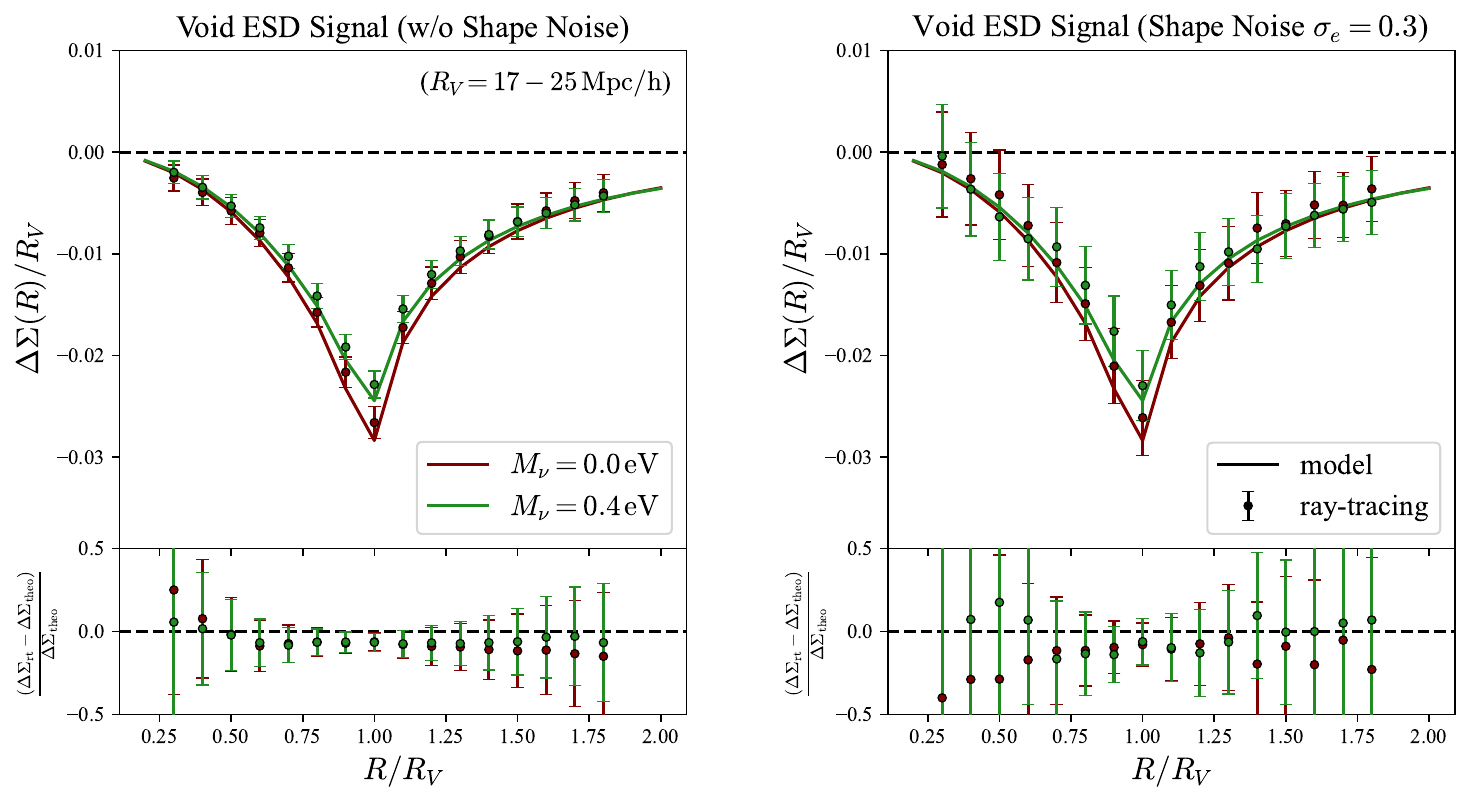}
\caption{Comparison between void lensing signal measured in ray-tracing mocks (data points) and theoretical models (solid lines). The radius ranges of measurement are $0.3 \le R/R_V \le 1.8$ with a bin size $\Delta R=0.1R_V$. Different colors indicate the measurements in simulations with $M_{\nu}=0.0\,{\rm eV}$ (red) and $M_{\nu}=0.4\,{\rm eV}$ (green). The left panel shows the measurements from mock source catalogues without shape noise, while the right panels display the results including shape noise $\sigma_e=0.3$. The errorbars are estimated with 15 realizations in Section~\ref{subsec:covariance}. Two bottom panels show the differences between the mock measurements ($\Delta \Sigma_{\rm rt}$) and models ($\Delta \Sigma_{\rm theo}$).}
\label{Fig_rt_ESD}
\end{figure*}

\subsection{Void Lensing Signal $\Delta \Sigma(R)$}
\label{subsec:void_esd}

As stated in Section~\ref{sec:lensing_formalism}, the differences in void density profiles leave imprints on the weak lensing signal. Following Eqs (\ref{eq:kappa_int}) and (\ref{eq:esd}), we theoretically model the weak lensing signal $\Delta \Sigma$ of voids from the measured 3D density profiles of total matter (both dark matter and neutrinos) presented in Section~\ref{subsec:void_profile}. The results are shown in Fig.~\ref{Fig_theo_ESD}. Due to the positive density slope at the interior of voids, $\Delta \Sigma(R)$ shows a minimum around the void boundary, which is consistent with previous void lensing studies and measurements from observational data \cite{ocvf_finder, Martin2025}. Similar to Fig.~\ref{Fig_vprof}, the statistical errors of $\Delta \Sigma$ are invisible as well (S/N ratio $\sim 10^3$) due to the large number of voids included in the calculation. The comparison between different neutrino masses shows that more massive neutrinos result in a less prominent void lensing signal (i.e. a higher $\Delta \Sigma(R)$ for large voids), consistent with the fact that nonzero neutrino mass results in more homogeneous matter distribution.

To validate the theoretical model shown in Fig.~\ref{Fig_theo_ESD}, we also measure the void lensing signals in the ray-tracing mocks presented in Section~\ref{subsec:raytracing}. The measurements are conducted in adaptive radius bins with resolution $0.1\,R_V$ for voids with different sizes. To simplify the analysis, we only consider the cross-correlation between two single redshift bins, and place all mock source galaxies at $z_s=0.75$. We investigate the cases both without and with shape noise ($\sigma_e=0.3$), with a source number density $10/{\rm arcmin}^2$. 

Fig.~\ref{Fig_rt_ESD} shows the measurements with theoretical models overlaid. For readability, here we only plot results for $M_{\nu}=0.0$ and $0.4\,{\rm eV}$. The errorbars of the data points are from the diagonal terms of the covariance matrix, the estimate of which is presented in Section~\ref{subsec:covariance}. A clear alignment between theoretical model and measurement from mock data can be found in the left panel, which is free of shape noise. With shape noise included, the lensing signals still follow the models well, although the errors increase significantly. The difference in $\Delta \Sigma$ between $M_{\nu}=0.0$ and $0.4\,{\rm eV}$ reaches $\sim1\,\sigma$ level in the vicinity of void boundary, in the case with shape noise, which indicates a potential constraint on neutrino mass from void lensing effect. We further investigate the constraining power in the next section.

\begin{figure*}
\centering
\includegraphics[width=\textwidth]{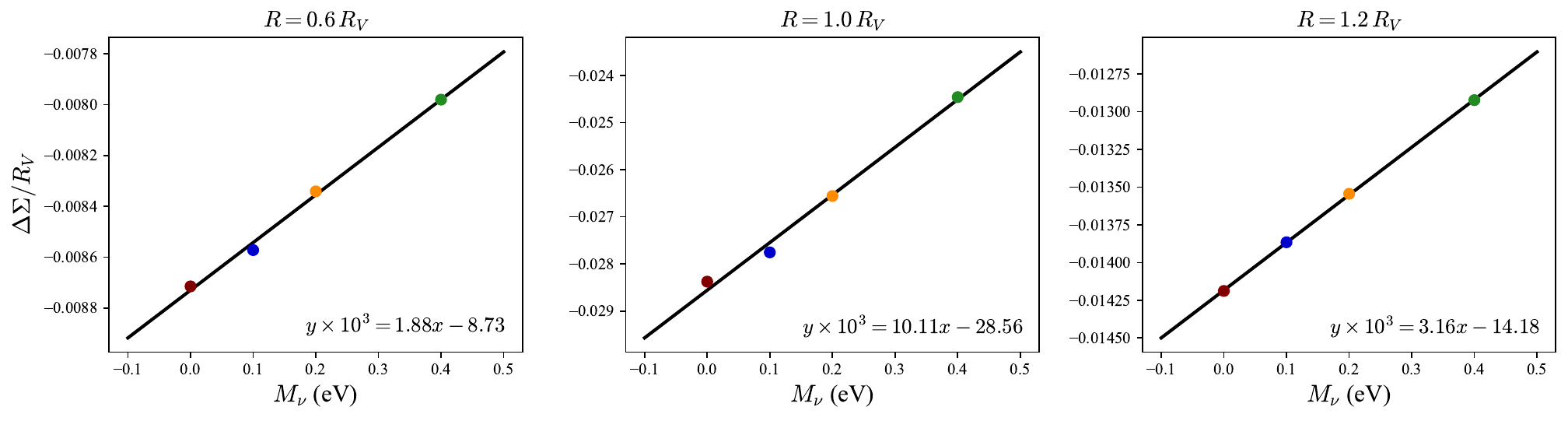}
\caption{Linear fitting of the relation between $\Delta\Sigma$ and $M_{\nu}$ at three different radii. The colors of data points are consistent with Fig.~\ref{Fig_vprof} and Fig.~\ref{Fig_theo_ESD}. The jackknife errorbars of data points are too small to be visible. The functions in each panel present the best-fit linear relations at different radii.}
\label{Fig_Mnu_linear}
\end{figure*}

\section{Constrain Neutrino Mass with Void Lensing Signal}
\label{sec:constraint}

As the comparison in Section~\ref{subsec:void_esd} validates the modelling of $\Delta \Sigma$ from void density profile, in this section, we further investigate the constraint on neutrino mass based on the theoretical models. We present the data vector and linear fitting of $\Delta\Sigma-M_{\nu}$ in Section~\ref{subsec:data_model}, and the estimate of covariance matrix in Section~\ref{subsec:covariance}. Finally we conclude with the sampling of posterior and discussions (Section~\ref{subsec:posterior}).

\begin{figure*}
\centering
\includegraphics[width=0.8\textwidth]{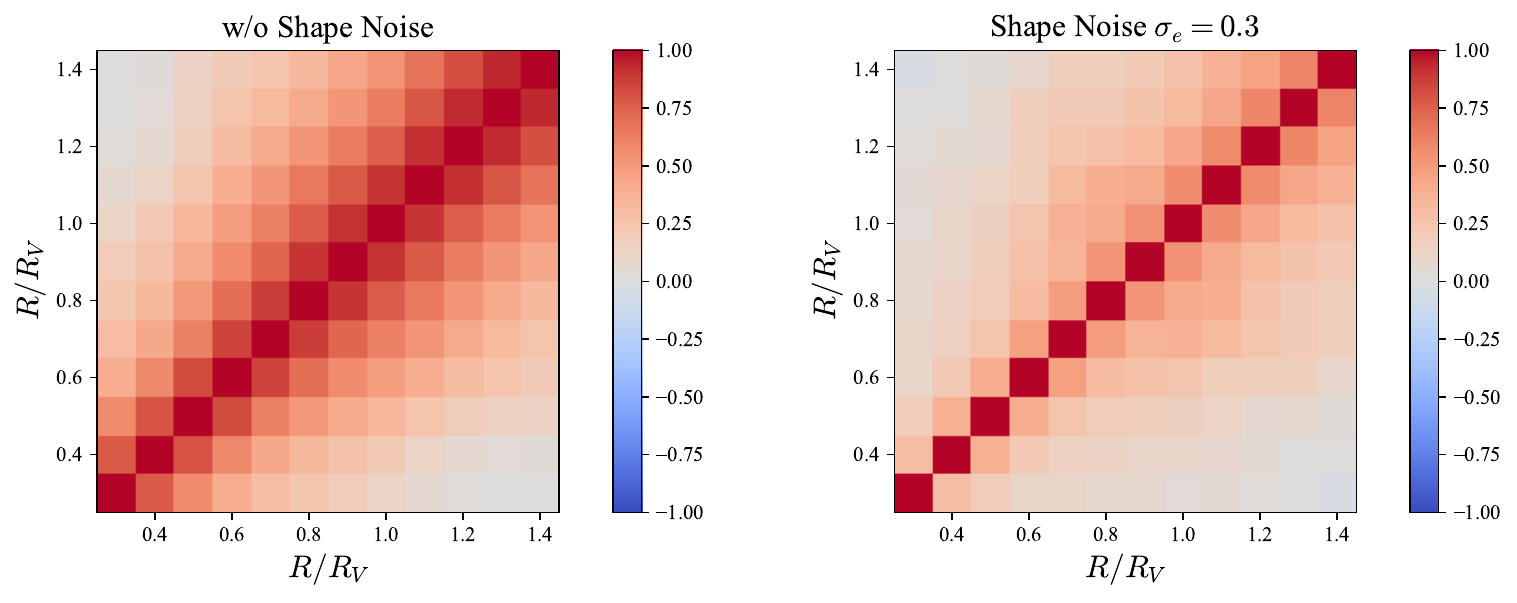}
\caption{Correlation matrix estimated with 15 realizations. The matrix is obtained by normalizing the covariance matrix with its diagonal elements. The two panels display results without (left) and with (right) shape noise.}
\label{Fig_corr_matrix}
\end{figure*}

\subsection{Data and Model}
\label{subsec:data_model}
Fig.~\ref{Fig_theo_ESD} shows that the differences in void lensing signal exist at all radii but become more prominent around void boundary. Therefore, we use data points at $0.3\le R/R_{V} \le 1.4$ for the following analysis.
Here we re-bin the theoretical $\Delta \Sigma(R)$ model into radius bins with an interval $R=0.1\,R_V$, which is same as the resolution of the measurements in ray-tracing mocks.
We then follow Ref.~\cite{Su2023} to combine all voids with weight $1/R_V$ and present void lensing signal as $\Delta \Sigma/R_V$ (we omit $R_V$ in the remaining text) in Figs.~\ref{Fig_theo_ESD} and~\ref{Fig_rt_ESD}. The investigation of other weights to optimize the S/N ratio of the measurement is out of the scope of this work, and we dedicate it to future work.

From the data vectors of different neutrino masses, we find that the relation between $\Delta \Sigma(R)$ and $M_{\nu}$ can be well fitted by linear functions, with slopes changing with radius $R$. We show the fitting at three different radii $R/R_V=0.6,\,1.0,\,1.2$ in Fig.~\ref{Fig_Mnu_linear}, and have checked the linear relation always persists in the radius range we use for inference. We note that the errorbars become invisible in the plot, as the jackknife errors for theoretical $\Delta \Sigma(R)$ model used here are extremely small (see Section~\ref{subsec:void_profile}), but we take the errors into account when doing the fitting. We then use the suite of best-fit linear relations at different radii as the model in the likelihood.

\subsection{Covariance Matrix}
\label{subsec:covariance}
Besides the model discussed in Section~\ref{subsec:data_model}, covariance is also necessary for inference. Limited by the number of simulations, we cannot follow the standard way to estimate covariance from a suite of simulations with different initial conditions. Therefore, we generate 15 realizations of ray-tracing mocks using box-remapping described in Section~\ref{subsec:raytracing} and follow the method presented in to derive the covariance matrix. Ref.~\cite{Escoffier2016} proposed an effective method to improve the covariance estimation from a limited number of realizations by combining jackknife resamplings in each realization, which can attain the same accuracy as using $\sim7$ times more realizations. The estimator $\bar{C}_{ij}$ presents as
\begin{equation}
    \bar{C}_{ij} = \frac{1}{N_M}\sum_{m=1}^{N_M}\hat{C}^{(m)}_{ij},
\end{equation}
\begin{equation}
    \hat{C}_{ij} = \frac{N_{JK}-1}{N_{JK}}\sum_{k=1}^{N_{JK}} (y_i^k-\bar{y}_i)(y_j^k-\bar{y}_j);
\end{equation}
where $N_{JK}$ is the number of jackknife subsamples in each realization and $N_M$ is the total number of realizations used. $\bar{y}_i$ is the data vector average over all jackknife replica in each realization. In this work, we adopt $N_M=15$ and $N_{JK}=64$. The correlation matrices obtained from normalizing the covariance matrix with diagonal terms are presented in Fig.~\ref{Fig_corr_matrix}, which are smooth and show marked correlations between neighboring bins.

\begin{figure}
\centering
\includegraphics[width=0.40\textwidth]{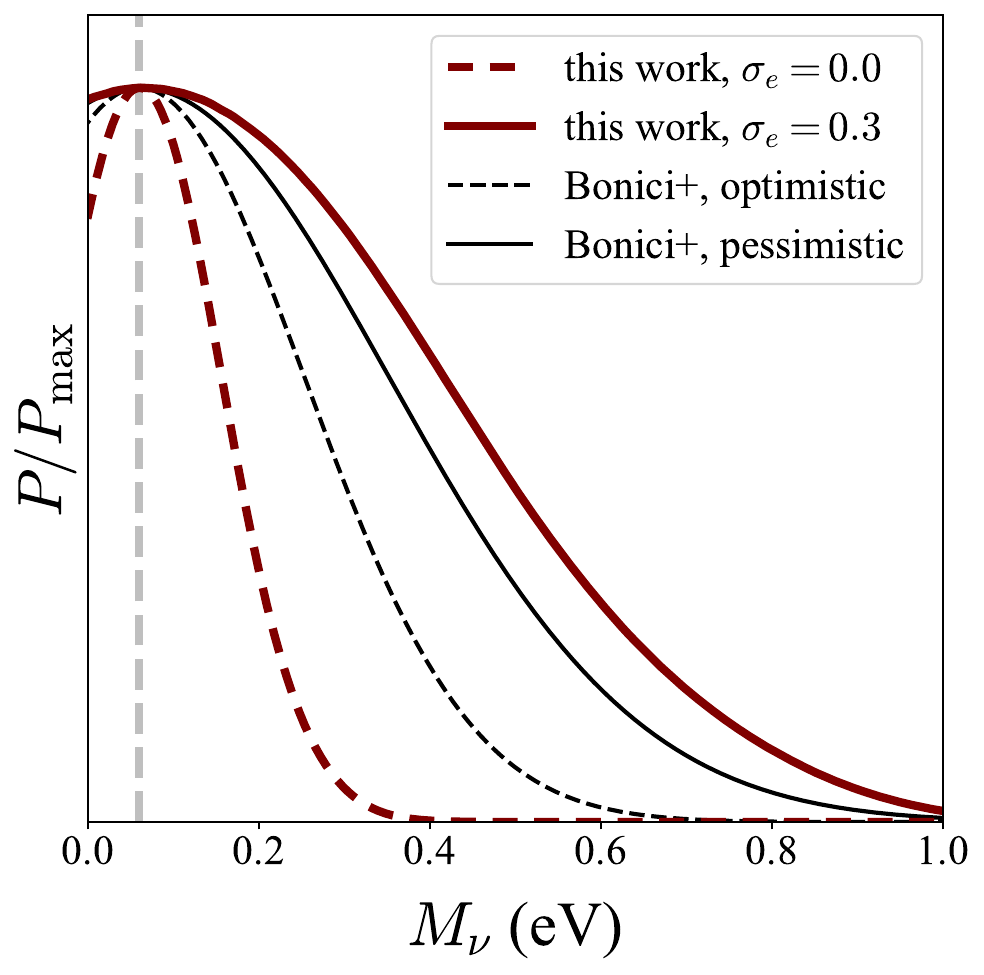}
\caption{Posteriors of neutrino mass in the cases without (red dashed line) and with (red solid line) shape noise. The vertical gray dashed line indicates the fiducial value $M_{\nu}=0.06\,{\rm eV}$. As a comparison, the black lines show the optimistic and pessimistic forecast results of Ref.~\cite{Euclid_neutrino} (see the Table 5 therein), which utilizes the large-scale cross-correlation between voids and shear catalogues and differs from our method. These two approaches are complementary and can be combined to yield improved constraints.}
\label{Fig_Mnu_posterior}
\end{figure}

It is noted that, as stated in Section~\ref{subsec:raytracing}, we apply several approximations when generating ray-tracing mocks. For instance, to utilize the entire slice cross-section at $z=0.3$, it is required to complement the slices at $z>0.3$ through the periodic boundary condition, which results in the repetition of path structures at $z>0.3$ in different realizations. In addition, each slice originates from the entire box of different snapshots in the same simulation, which also leads to the appearance of similar structures in different slices. Both of these approximate operations may lead to slightly inaccuracies in covariance. However, as shown in Fig.~\ref{Fig_corr_matrix}, the covariance gets dominated by shape noise when taking it into account. Due to the fact that the constraint on neutrino mass given by void lensing is still quite loose (Section~\ref{subsec:posterior}), we do not expect that these limitations would bias our final results significantly, and leave more careful covariance estimation to future work.

\subsection{Posterior on $M_{\nu}$}
\label{subsec:posterior}
With the linear model and covariance described above, we sample the posterior of neutrino mass with {\sc ultranest} \cite{ultranest}, assuming a Gaussian likelihood and a fiducial cosmology with neutrino mass $M_{\nu}=0.06\,{\rm eV}$. The posteriors are shown in Fig.~\ref{Fig_Mnu_posterior} with a truncation to $M_{\nu}>0$. With our setup (see Section~\ref{subsec:raytracing}), the void lensing effect can independently provide a constraint on neutrino mass as $\sigma(M_{\nu})=0.096\,{\rm eV}$ ($M_{\nu}<0.232\,{\rm eV}$, 95\% C.L.) in the absence of shape noise, and it becomes $\sigma(M_{\nu})=0.340\,{\rm eV}$ ($M_{\nu}<0.707\,{\rm eV}$, 95\% C.L.) when including a $\sigma_e=0.3$ shape noise.

Ref.~\cite{Euclid_neutrino} also investigates the cross-correlation between cosmic voids and weak lensing, but presents forecasts based on a \textit{Euclid}-like setup with a sky area of $\sim 15000\,{\rm deg}^2$, a source number density of $3/{\rm arcmin}^{2}$ and shape noise of $\sigma_e=0.3$, which slightly differs from ours. We also include their results in Fig.~\ref{Fig_Mnu_posterior} for comparison. Their analysis combines the auto-clustering of weak lensing shear and voids with their cross-correlation, and obtains constraints of $\sigma(M_{\nu})=0.193\,{\rm eV}$ ($0.292\,{\rm eV}$) in the optimistic (pessimistic) scenarios for different modelling of void bias, which are at a similar level to our results. However, this work focused on large-scale cross-correlation only to avoid explicit modelling of void density profile, which is actually complementary to our method. Given that they identify 2D voids from photometric galaxy catalogues, this approach also merits further investigation for DT voids in the future.

It should be noted that our current result is limited to only one parameter $M_{\nu}$, while Ref.~\cite{Euclid_neutrino} constrain multiple cosmological parameters simultaneously. However, as stated in Ref.~\cite{Massara2015}, void lensing would potentially help break the degeneracy between neutrino mass and other cosmological parameters, such as $\Omega_m$ and $\sigma_8$. Therefore, we anticipate a tighter cosmological constraint from the synergy between void lensing and other statistics, such as galaxy lensing ‘$\rm 3\times2pt$’. In addition, tomographic analyses combining voids across multiple redshift bins adopted in Ref.\cite{Euclid_neutrino} also help to improve the constraining power. These extensions merit further investigation in future work.

\section{Discussions and Summary}
\label{sec:summary}

Massive neutrinos leave distinct imprints on both the cosmic expansion history and evolution of LSS, providing important opportunities to constrain their total mass with cosmological observations \cite{particle_physics_review}. 
As the underdense regions in the Universe, cosmic voids offer essential environments for investigating neutrino properties \cite{Massara2015}.

In this work, we investigate the constraining power of weak lensing effect of cosmic voids on neutrino mass, using realistic mocks generated from $N$-body simulation with varying total neutrino masses (0 eV, 0.1 eV, 0.2 eV and 0.4 eV). For the void catalogues, we fit an HOD model to BOSS LOWZ galaxies in $0.2<z<0.4$, identify DT voids with the {\sc dive} void finder developed by Ref.~\cite{Zhao2016} and select voids with $17<R_V<25$\,Mpc/h.  The weak lensing shear catalogues are generated through multiplane ray-tracing simulations \cite{Hilbert2009}. We adopt a sky area of $\sim 8400\,{\rm deg}^2$ and a lensing source number density of $10/{\rm arcmin}^{-2}$, which are comparable to next-generation surveys. For simplicity, we place all source galaxies at $z_s=0.75$ and fix cosmological parameters other than $M_{\nu}$.

Our main findings are summarized as follows:
\begin{enumerate}
    \item The density profiles of DT voids show clear sensitivity to the total neutrino mass $M_\nu$. The presence of massive neutrinos leads to higher void interior densities and reduced density peaks at the void boundary (Fig.~\ref{Fig_vprof}). These effects become more prominent as with neutrino mass increasing.

    \item Variations of void density profiles leave observable imprints in the weak lensing signal. For voids of the same size, massive neutrinos result in a less pronounced lensing signal (Fig.~\ref{Fig_theo_ESD}). In particular, the excess surface density profile, $\Delta\Sigma(R)$, exhibits a clear linear dependence on the neutrino mass within $R<1.4\,R_V$ (Fig.~\ref{Fig_Mnu_linear}).

    \item The weak lensing signal of DT voids yields an independent constraint on the total neutrino mass of $\sigma(M_\nu)=0.096\,{\rm eV}$, in the absence of shape noise, and $\sigma(M_\nu)=0.340\,{\rm eV}$ when including realistic galaxy shape noise with $\sigma_e=0.3$. These constraints are comparable to those obtained in other weak lensing studies. 

    \item The void density profile provides an effective approach for modelling the void lensing signal, as validated across different cosmologies (Fig.~\ref{Fig_theo_ESD}) and void size bins (Appendix~\ref{appendix_a}). This alleviates the need for full ray-tracing simulations and facilitates the development of emulator-based models for void lensing in future work.
    
\end{enumerate}

Looking ahead, there are several aspects to be investigated in future works. Firstly, a more delicate modelling of the void lensing signal and accurate estimation of the covariance are required for future applications to observational data. Ref.~\cite{Su2025} investigated the simulation-based inference (SBI) for void lensing modelling, while emulator-based approaches have also significantly facilitated cosmological analyses (e.g. Refs.~\cite{Aric2021, Mancini2022}). This also necessitates higher-fidelity mocks that incorporate more observational effects beyond the shape noise, such as imaging systematics, magnification bias and intrinsic alignments \cite{Lange2024}.

Other effects not included in this work, such as baryonic feedback and RSD, might also modify void density profiles. Several works (e.g. Refs.~\cite{Paillas2017, Schuster2024}) have shown that baryonic effects have a minimal impact on the density profiles of watershed voids, and we therefore anticipate a similar result for DT voids. Nevertheless, voids identified in redshift space tend to show anisotropic density profiles \cite{Pisani2014, rsd_void_xcorr}, which leads to variations in the lensing signal. A possible approach to mitigate the RSD effect is to identify voids in the reconstructed galaxy catalogues, which has been investigated for measurements of the void-galaxy correlation function \cite{Nadathur2019, radinovic2023, Degni2026}. We leave a detailed discussion of these effects for future work.

Furthermore, constraints on neutrino mass are subject to degeneracies with other parameters (such as $\Omega_m$ and $\sigma_8$ \cite{Ivanov2020, Elbers2025}), which may lead to an overestimation of the constraining power on $M_{\nu}$, as all other $\Lambda$CDM parameters fixed in this work. As voids provide information complementary to overdense regions, combining void lensing with other statistics, such as void size function and weak lensing ‘3x2pt’ analysis, will break parameter degeneracies and yield tighter cosmological constraints. These synergies also facilitate our future extensions of this framework to more cosmological parameters beyond $M_\nu$.
In addition, analogous to galaxy-galaxy lensing, tomographic analyses of void lensing have the potential to enhance the constraining power, owing to both the redshift evolution of void properties and the increased void sample size \cite{Euclid_neutrino}.

In conclusion, our investigations based on mock data demonstrate that void lensing effect provides an effective and complementary probe for constraining neutrino mass with cosmological surveys. We anticipate that void lensing will help promote the cosmological constraints in existing cosmological surveys, including DESI, \textit{Euclid} \cite{Euclid_overview}, KiDs, DES, HSC, as well as future programs, such as MUST \cite{MUST_paper}, CSST \cite{CSST_overview} and LSST \cite{lsst_paper}.

\begin{figure*}
\centering
\includegraphics[width=\textwidth]{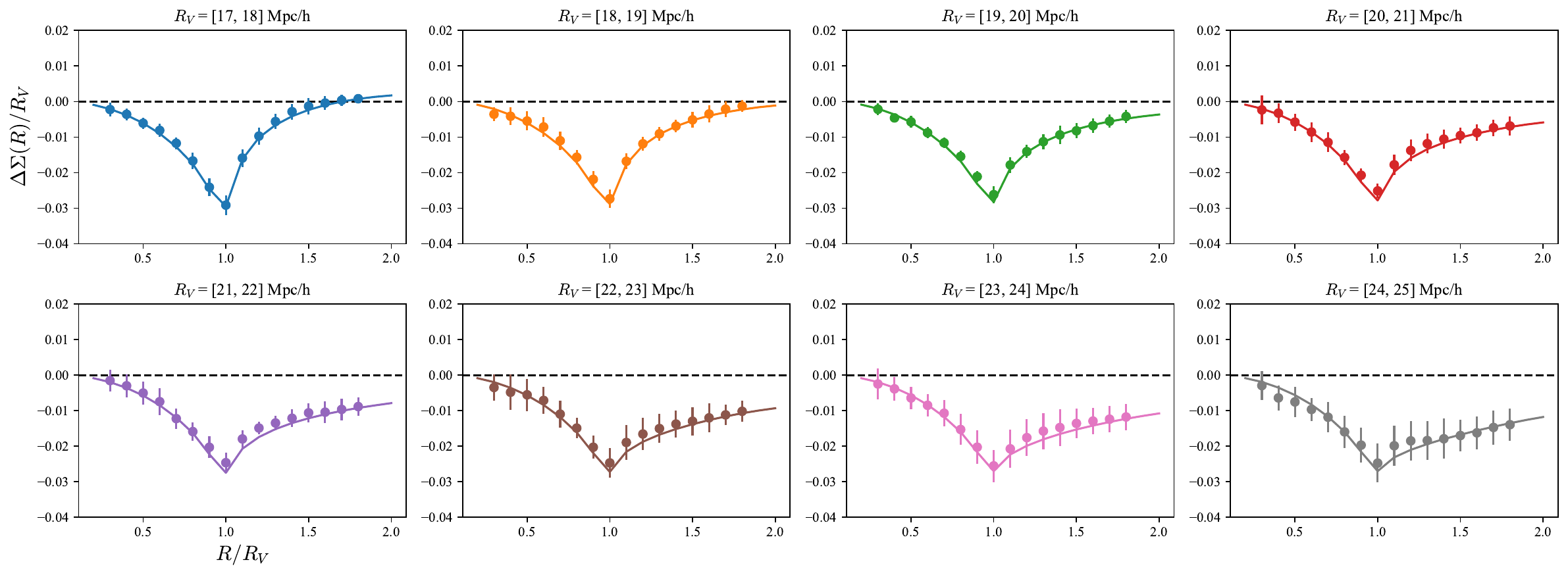}
\caption{Comparison between the theoretical models of void lensing signal $\Delta \Sigma$ (solid lines) and measurements from ray-tracing mocks (data points) in different radius bins. The results shown here are from the massless-neutrino cosmology.}
\label{Fig_ESD_bin_wo_noise}
\end{figure*}

\begin{figure*}
\centering
\includegraphics[width=\textwidth]{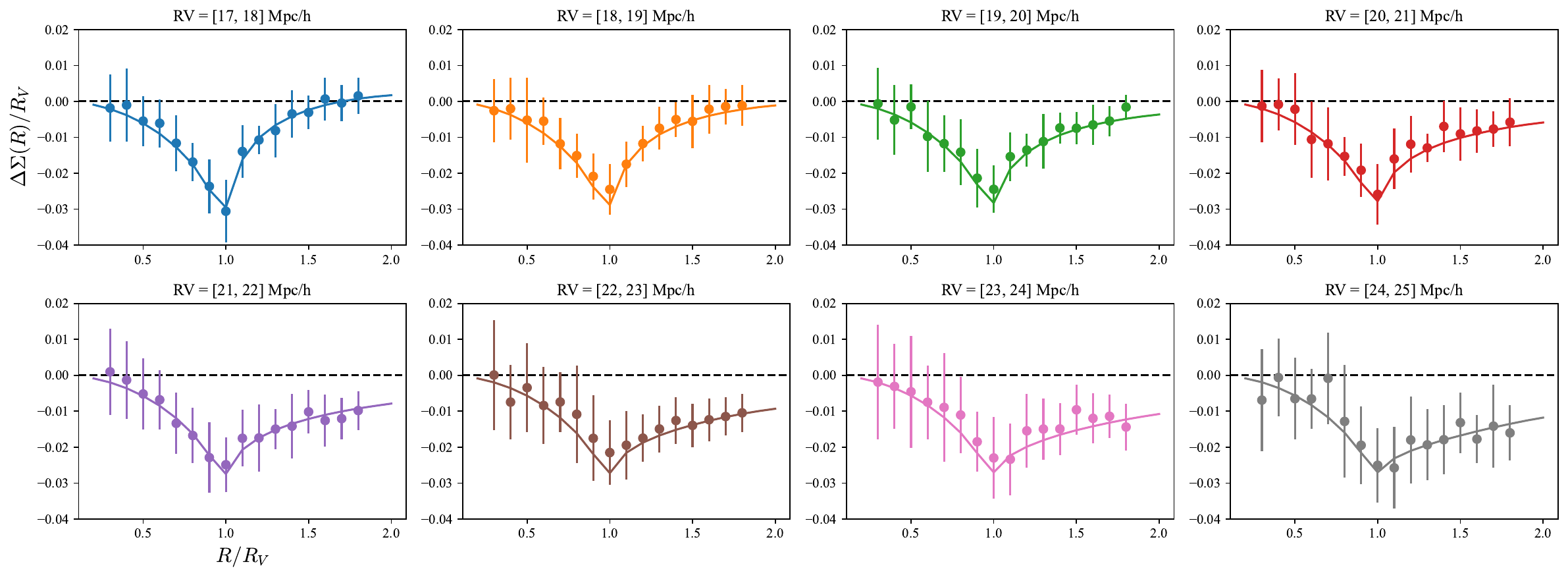}
\caption{Same as Fig.~\ref{Fig_ESD_bin_wo_noise}, with shape noise $\sigma_e=0.3$ included.}
\label{Fig_ESD_bin_w_noise}
\end{figure*}

\begin{acknowledgments}

We would like to thank Qiao Wang for helpful discussions on the ray-tracing code. WSX appreciates the insightful discussions with Siyi Zhao and Eric Jullo. We acknowledge the support from National Key R\&D Program of China (Grant No. 2023YFA1605600) and National Natural Science Foundation of China (Grant No. 12303005). YL acknowledges the support from Shuimu Tsinghua Scholar Program (No. 2022SM173) and Swiss National Science Foundation research grant "Cosmology with 3D Maps of the Universe" (No. 200020\_207379).

WSX acknowledges the Tsinghua Astrophysics High-Performance Computing platform for providing computational and data storage resources that have contributed to the research results within this paper.

In this work, we also use the following python packages in addition to those already cited in the main content: {\sc astropy} \cite{astropy}, {\sc numpy} \cite{numpy}, {\sc matplotlib} \cite{matplotlib}, {\sc scipy} \cite{scipy} and {\sc getdist} \cite{getdist}.

\end{acknowledgments}

\begin{appendix}

\section{Void Lensing Signal in Different Radius Bins}
\label{appendix_a}

Figs.~\ref{Fig_ESD_bin_wo_noise} and~\ref{Fig_ESD_bin_w_noise} present the measurements of void lensing signal $\Delta \Sigma(R)$ in different radius bins, along with the corresponding theoretical models derived from void density profiles, in the simulation with $M_{\nu}=0.0\, {\rm eV}$. We find great agreements between models and measurements across all radius bins, which further validate the modelling method. The errorbars increase for larger voids, owing to the much lower number density of voids with large radius. 

The results in other cosmologies are similar, and the trend of a weaker void lensing signal with increasing neutrino mass persists in all radius bins.

\section{Void Size Function}
\label{appendix_b}

\begin{figure}[htbp]
\centering
\includegraphics[width=\linewidth]{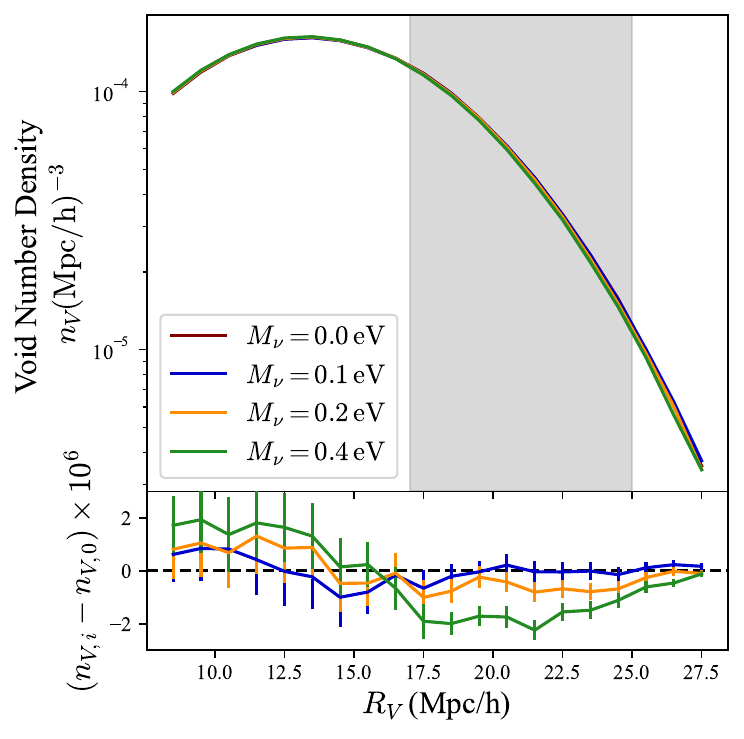}
\caption{Void size function measured in simulations with different neutrino masses. The radius bin used here is $\Delta R_V=1\,$Mpc/h. The shaded region in the upper panel indicates the void radius range that we use in this work. The bottom panel displays the differences between results of massive neutrinos and the massless-neutrino case.}
\label{Fig_VSF}
\end{figure}

We present the void size function (VSF) measured in simulations with different neutrino masses in Fig.~\ref{Fig_VSF}. The results demonstrate the sensitivity of the VSF to neutrino mass, which results in an excess of small voids and a reduced amount of large voids with the existence of massive neutrinos. Our results are consistent with Ref.~\cite{Massara2015}, but show a lower statistical significance. We attribute this to the potential correlation between the VSF of DT voids and galaxy clustering, as we control the 2-point clustering of our galaxy samples with HOD fitting.

\section{Ray-tracing Slices from Box Remapping}
\label{appendix_c}

Following Ref.~\cite{Carlson2010}, for a simulation box with periodic boundary conditions, the cuboid remapping method can transform a cubic box to a rectangular cuboid with preserving the continuity and avoiding repetition of structures. 
In this work, we adopt the method on the $x$-$z$ plane to transform our simulation box of side length $L=1500\, {\rm Mpc/h}$ into a cuboid of $(1500\sqrt{8^2+9^2},\, 1500,\, 1500/\sqrt{8^2+9^2}) \, {\rm Mpc/h}$ (see Fig.2 in Ref.~\cite{Carlson2010} for an illustration). We construct the ray-tracing lightcones by stacking a sequence of remapped boxes up to $z_s = 0.75$ along their shortest side. This results a suite of lensing ‘slices’ with thickness of $1500/\sqrt{8^2+9^2}\approx 125\, {\rm Mpc/h}$ (except for the first and last slices), which is close to the safe range presented in Ref.~\cite{Zorrilla2020}. Each single slice therefore comes from a full remapped box, while the inside structures are taken from the closest simulation snapshots.

In the analysis, we only use voids at $z_l=0.3$ for the void lensing cross-correlation. For this purpose, we make use of the full remapped slice at $z_l=0.3$, which allows for a split into 15 subregions, each covering $8400\, {\rm deg}^2$. As there is no structure repetition within each remapped slice, the same void does not appear multiple times in different subregions.

However, the ray-tracing algorithm requires all slices to subtend at least the same angular area. For slices at $z>0.3$, the corresponding physical area therefore exceeds a single remapped slice. We complement these additional areas using the periodic boundary conditions in the two transverse dimensions, which are preserved by the remapping method. 

This procedure actually leads to the repetition of the same line-of-sight structures across the 15 subregions, and may affect the covariance estimation \cite{Vecchi2025}. Nevertheless, this would not bias our results, as the covariance will be dominated by galaxy shape noise as shown in Section~\ref{subsec:covariance}.

\begin{figure*}[htbp]
\centering
\includegraphics[width=\textwidth]{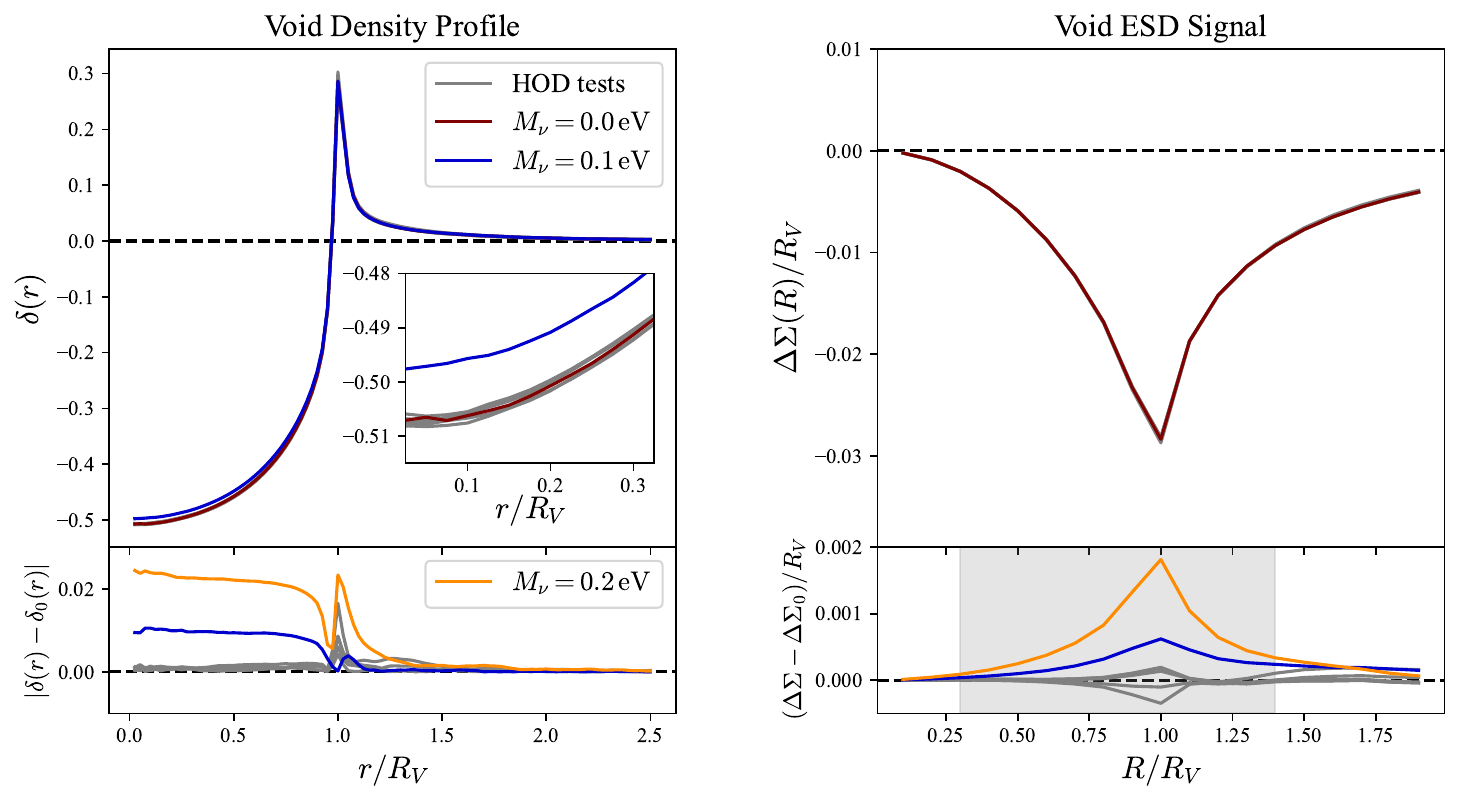}
\caption{Void density profiles and lensing signals for five additional sets of HOD parameters (gray solid lines). The results for $M_{\nu}=0.0,\,0.1,\,0.2\,{\rm eV}$ are overlaid for comparison. The shaded region in the right bottom panel indicates the radius range of $\Delta\Sigma(R)$ used for the analysis in Section~\ref{sec:constraint}.}
\label{Fig_test_HOD}
\end{figure*}

\section{HOD Uncertainties}
\label{appendix_d}

In this section, we investigate how uncertainties from the HOD fitting propagate into the modelling of the void density profile. We take the simulation with massless neutrino as an example, select five additional sets of HOD parameters around the best-fit values, and compute the combined void density profiles and lensing signals for these different parameter sets using the same pipeline. We present the results in Fig.~\ref{Fig_test_HOD}.

From Fig.~\ref{Fig_test_HOD}, we find that, different HOD parameters yield very similar density profiles, with the most obvious uncertainties showing up around the void radius. Given that most of the constraint on neutrino mass originates from the inner underdense regions, the variations are much smaller than the differences between the $M_{\nu}=0.0\,{\rm eV}$ and $M_{\nu}=0.1\,{\rm eV}$ cosmologies. The right panel further demonstrates that the HOD uncertainties are less prominent in the radius range that we use for analysis in this work (the gray shaded region). Therefore, we conclude that HOD fitting would not bias our main conclusions.

We attribute the relatively large uncertainties around the void radius to the scale range used in the HOD fitting. Specifically, in this work we only use the projected two-point correlation function $w_p(r_p)$ at scales larger than $\sim5\,{\rm Mpc/h}$ (see Fig.\ref{Fig_HOD}), which limits the ability to constrain the small-scale galaxy distribution, and leads to increased noise at the overdense regions. Nevertheless, the uncertainties remain smaller than the differences between $M_{\nu}=0.0\,{\rm eV}$ and $M_{\nu}=0.2\,{\rm eV}$ cosmologies. We leave a more sophisticated treatment of HOD fitting, for instance using machine learning based emulators \cite{Zhai2019}, to future work.

\end{appendix}

\bibliography{reference}

\end{document}